\documentclass[10pt,twocolumn,letterpaper]{article}

\usepackage{cvpr}              
\definecolor{cvprblue}{rgb}{0.21,0.49,0.74}
\usepackage[pagebackref,breaklinks,colorlinks,allcolors=cvprblue]{hyperref}

\def\paperID{19} 
\def\confName{CVPR}
\def\confYear{2026}

\title{RAG4Outcome: A Retrieval-Augmented Multimodal Framework for \\ Prognostic Prediction in Chronic Osteomyelitis}

\author{
Daqian Shi\\
Queen Mary University of London\\
London, UK\\
{\tt\small d.shi@qmul.ac.uk}
\and
Pei Han\\
Shanghai Sixth People’s Hospital \\Affiliated to SJTU School of Medicine\\
Shanghai, CN\\
{\tt\small hanpei\_cn@163.com}
\and
Jishizhan Chen\\
University College London\\
London, UK\\
{\tt\small jishizhan.chen@ucl.ac.uk}
\and
Yang Wang\\
Shanghai Sixth People’s Hospital\\
Shanghai, CN\\
{\tt\small hongchayang@sina.com}
\and
Xiaolei Diao$^{*}$\\
University College London\\
London, UK\\
{\tt\small xiaolei.diao@ucl.ac.uk}
\and
Xianyou Zheng$^{*}$\\
Shanghai Sixth People’s Hospital \\Affiliated to SJTU School of Medicine\\
Shanghai, CN\\
{\tt\small zhengxianyou@situ.edu.cn}
\and
Pengfei Cheng$^{*}$\\
Shanghai Sixth People’s Hospital \\Affiliated to SJTU School of Medicine\\
Shanghai, CN\\
{\tt\small chengpf@alumni.sjtu.edu.cn}
}

\begin{document}
\maketitle

\begingroup
\renewcommand\thefootnote{\fnsymbol{footnote}}
\footnotetext[1]{Corresponding author.}
\endgroup

\begin{abstract}
Chronic osteomyelitis presents substantial prognostic challenges due to its high recurrence risk and complex postoperative recovery trajectories. Traditional assessment often relies on manual scoring systems, which limit scalability, efficiency, and consistency in clinical practice. Furthermore, the heterogeneous nature of clinical data poses challenges for current multimodal learning approaches that require aligned inputs and large annotated datasets. In this work, we propose RAG4Outcome, a retrieval-augmented generation (RAG) framework for prognostic prediction in chronic osteomyelitis. Our method integrates multimodal clinical data, including PET-CT imaging reports, structured surgical and diagnostic records, and unstructured follow-up notes, into a unified prediction pipeline. By combining a domain-specific retrieval corpus with expert-guided prompting, the framework enables more interpretable, evidence-grounded, and clinically reliable prognosis. Preliminary results on real-world cases demonstrate promising effectiveness and clinical alignment, highlighting the potential of RAG4Outcome for AI-assisted infection management and postoperative decision support.
\end{abstract}

\section{Introduction}
\label{sec:Introduction}
Chronic osteomyelitis is a highly challenging orthopedic infection characterized by high recurrence risk and complex postoperative recovery trajectories\cite{ucckay2012chronic, panteli2016chronic, schmitt2017osteomyelitis}. Accurate prognostic prediction is critical for tailoring personalized treatment strategies and ensuring long-term care \cite{dudareva2021systematic}. However, current clinical assessment methods typically rely on expert-defined scoring systems and manual evaluation processes \cite{cierny1985clinical}, which are labor-intensive, time-consuming, and difficult to scale across different care settings. This limits their effectiveness for continuous monitoring and early risk stratification in real-world scenarios.

In practice, clinicians must reason over a broad range of heterogeneous data types \cite{lew2004osteomyelitis, wu2023knowlab, wu2024slava}, including medical imaging reports, structured Electronic Health Records (EHR), and unstructured reports, to assess a patient’s recovery and recurrence risk. These multimodal data sources are often incomplete, asynchronous, and loosely aligned, posing significant challenges for conventional multimodal learning models\cite{xia2024mmed, amugongo2025retrieval}. Existing AI-based methods for postoperative prediction often rely on rigid data alignment, require large amounts of well-labeled multimodal data\cite{amugongo2025retrieval}, and tend to generalize poorly across real-world clinical environments\cite{firat2025management, ke2025retrieval}. 
Moreover, critical indicators such as infection type, surgical timing, metabolic biomarkers, and symptom evolution are often embedded in unstructured clinical texts, making them difficult to extract and interpret.


To address the above challenges, we propose RAG4Outcome, a retrieval-augmented multimodal framework designed for prognostic prediction in chronic osteomyelitis. Our method integrates multiple forms of clinical data into a unified prediction model, including PET-CT \cite{kapoor2004introduction} imaging reports, structured EHRs, diagnostic and surgical reports, and follow-up documentations. A central Retrieval-Augmented Generation (RAG) module integrates external medical knowledge retrieved from a domain-specific corpus to produce evidence-grounded outcome predictions. This design enables reliable generation without requiring strict data alignment or exhaustive annotations \cite{xiong2024benchmarking, amugongo2025retrieval, ke2025retrieval, ng2025rag}.
To enhance clinical interpretability, we incorporate 12 prognostic indicators identified by orthopedic experts as the most relevant to recovery outcomes in chronic osteomyelitis. These factors guide the structured prompt construction and support targeted retrieval and reasoning. We validate our approach using a real-world dataset\footnote{The dataset was collected under institutional ethical approval, ensuring patient privacy and data governance compliance.} from anonymized chronic osteomyelitis patients treated at a tertiary care center, each with follow-up in 3–6 years.  Case studies on representative patients demonstrate that RAG4Outcome achieves high consistency with clinical scoring systems, while providing transparent, evidence-supported rationales. This work provides a promising step toward the development of reliable, interpretable, and scalable AI tools for infection prognosis and surgical decision support. 
Rather than replacing established clinical scoring systems, our goal is to complement them with a transparent retrieval-supported framework that can synthesize heterogeneous evidence and assist clinicians in real-world postoperative assessment.
Our contributions are summarized as follows:
\begin{itemize}
 \item We propose RAG4Outcome, a retrieval-augmented multimodal framework for clinical prognostic prediction in osteomyelitis, capable of integrating heterogeneous and partially missing clinical data into a unified model.
\item We design an interpretable prognostic framework that leverages twelve expert-defined indicators and a curated medical retrieval corpus to extract relevant clinical cues and construct structured prompts from multimodal inputs.
\item We conduct case-level evaluations using real-world patient data, demonstrating strong agreement with established scoring systems and the clinical utility.
 \end{itemize}

\section{Related Work}
\label{sec:Related Work}

\subsection{Prognostic Assessment in Osteomyelitis}

Prognostic assessment in chronic osteomyelitis has traditionally relied on structured scoring systems designed for orthopedic and infection-related outcome monitoring. One of the most widely used scales is the Lower Extremity Functional Scale (LEFS), a patient-reported outcome measure that evaluates lower-limb functionality \cite{binkley1999lower, dingemans2017normative, mehta2016measurement}. The Enneking system assesses musculoskeletal tumor patients' functional status and has been adapted for post-treatment outcome assessment \cite{wada2007construct}. The Cierny–Mader system evaluates chronic osteomyelitis cases based on anatomic type and host physiology, and is extensively used in clinical risk stratification \cite{panteli2016chronic, conway2021immunological}.  Different learning strategies are also introduced in the community to optimize the performance of deep learning-based methods \cite{shi2025competitive, shi2022charformer, chen2025minuscule, chang2025piezoelectric}. While clinically valuable, these systems require manual interpretation and exhibit subjectivity, limiting scalability and real-time application. Furthermore, studies suggest that current scoring methods fall short in supporting automated or continuous prognostic monitoring \cite{panteli2016chronic}, highlighting the need for AI-driven, interpretable alternatives.

\subsection{Language Models and Retrieval‑Augmented Generation in Medical AI}

Large Language Models (LLMs) have shown promise in medical text understanding, summarization, and clinical decision support. BioGPT \cite{luo2022biogpt} and Med-PaLM \cite{singhal2023large, singhal2025toward} demonstrate strong performance on biomedical QA and benchmark challenges. 
However, LLMs are prone to factual errors, commonly referred to as hallucinations \cite{zhang2025llm}, particularly in scenarios lacking reliable contextual information. To mitigate this, a hybrid architecture, called Retrieval-Augmented Generation, is introduced that combines retrieval-based evidence with generative capabilities. In clinical settings, RAG has been shown to improve factual consistency, support the interpretation of medical guidelines, and reduce hallucination risks \cite{xiong2024benchmarking, hasan2025llm, wang2025automated, shi2025graph}. Some of them introduce knowledge bases into the systems \cite{shi2020learning, shi2024kae}. Yang et al. \cite{ke2025retrieval} applied RAG for the interpretation of the guideline and demonstrated improvements in the factual correctness and the agreement of the clinician. Xiong et al. \cite{xiong2024benchmarking} established benchmarks for RAG in the medical domain and showed its superiority over closed-book LLMs in evidence-aware reasoning tasks. Recent applications include MedRAG for evidence-grounded diagnostic assistance \cite{hasan2025llm}, and a zero-shot RAG-based framework for automatic disease phenotyping from electronic health records, demonstrated on pulmonary hypertension as a case study \cite{thompson2023large}. By explicitly linking model outputs to retrieved evidence, RAG enhances response consistency and credibility, while also allowing flexible integration of heterogeneous data sources without requiring input alignment or model retraining.

\section{The Proposed Method}
\label{sec:Method}

\begin{figure*}[t]
    \centering
    \includegraphics[width=0.8\linewidth]{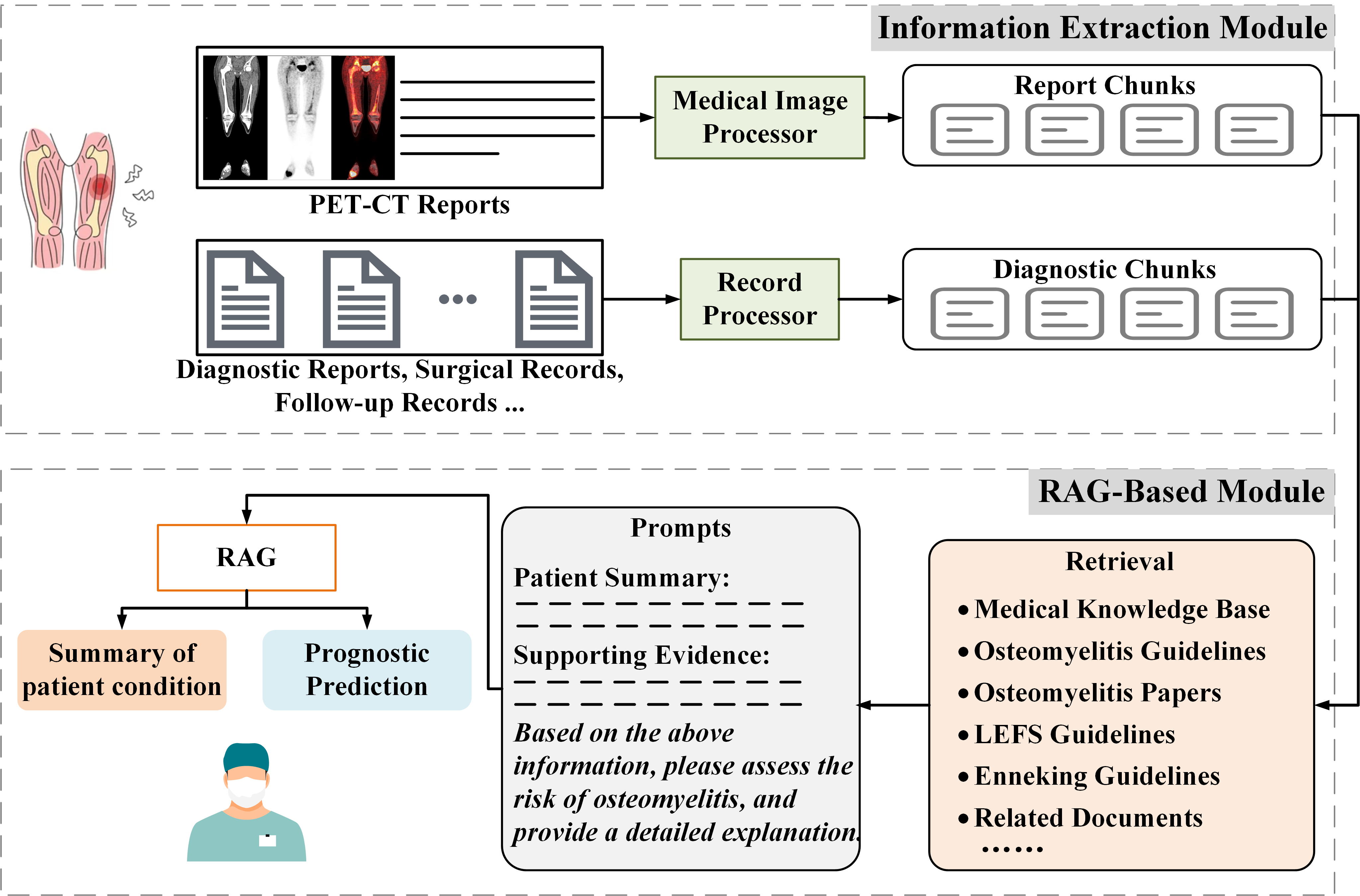}
    \caption{Overview of the RAG4Outcome framework. The system integrates PET-CT reports, EHR surgical records, and follow-up documentations through an information extraction pipeline, and then applies an RAG-based model to perform patient-level outcome prediction.}
    \label{fig:overall_structure}
\end{figure*}

In this section, we present the RAG4Outcome framework, a retrieval-augmented multimodal system designed to support clinical prognostic prediction in chronic osteomyelitis. As illustrated in Figure~\ref{fig:overall_structure}, the proposed pipeline consists of two primary modules: (1) an Information Extraction Module that processes multimodal clinical data into interpretable textual representations, and (2) a RAG-based Module that performs evidence-grounded reasoning for outcome prediction using large language models enhanced with domain-specific retrieval.

\subsection{Overview of The Framework}

Given a chronic osteomyelitis patient, the proposed framework takes three types of multimodal clinical data as input, including PET-CT imaging reports, structured EHRs such as diagnostic and surgical records, and unstructured follow-up documentation. These heterogeneous data sources are processed by modality-specific components, the Medical Image Processor and the Record Processor, to generate coherent and interpretable textual chunks, each representing key clinical factors defined by chronic osteomyelitis experts.

The extracted chunks are used to construct patient-specific prompts, which are subsequently passed to the RAG-base Module. During inference, RAG retrieves relevant domain-specific evidence from an external medical corpus and generates both a structured summary of the patient’s condition and a prognostic prediction. The final output includes a structured summary of the patient's condition and a prognostic prediction describing recovery outcome (including excellent, good, fair, and poor), along with supporting explanations grounded in both patient data and retrieved external references. Unlike prior approaches that require tightly aligned multimodal inputs, RAG4Outcome supports heterogeneous, asynchronous, and partially missing data without the need for modality alignment. This is achieved by independently processing each input source and transforming it into a standardized semantic representation suitable for retrieval and reasoning.


\subsection{Information Extraction Module}
We denote a patient case as a collection of heterogeneous clinical documents $\mathcal{D} = \{d_1, d_2, \dots, d_N\}$, where each $d_i$ corresponds to one of the following modalities: PET-CT imaging reports, structured surgical or diagnostic EHRs, and unstructured follow-up documentations. Each input type is processed by a modality-specific preprocessor, namely Medical Image Processor or Record Processor, to extract semantically meaningful representations for downstream retrieval and generation.

\noindent \textbf{Medical Image Processor.} PET-CT imaging reports are parsed into structured textual fields that describe radiological findings and anatomical annotations. We incorporate Qwen2.5-VL-3B to extract expert-defined indicators from the PET-CT image and its natural language reports. 

\noindent \textbf{Record Processor.} Surgical, diagnostic, and follow-up documents are segmented into sentences and sections. We apply Qwen3-4B as the processor to generate embeddings for sentences associated with predefined prognostic indicators, allowing targeted semantic understanding. 

Each document chunk $d_i$ is encoded into a dense vector representation via a pretrained domain-specific encoder:
\begin{equation}
    \mathbf{h}_i = \text{Processor}(d_i)
\end{equation}
where $\mathbf{h}_i \in \mathbb{R}^d$ is the resulting embedding of the document chunk $d_i$. Note that, missing modalities are handled with padding with null vectors $\mathbf{0}$ for structurally expected but unavailable modalities.


\subsection{RAG-based Module}

The RAG-based module serves as the core reasoning engine in our framework, integrating structured patient data and external clinical knowledge to generate prognostic predictions with explanatory support. This module consists of three key components: a domain-specific retriever, an expert-guided prognostic schema, and a retrieval-augmented generation model.

\noindent \textbf{Domain-Specific Retriever:}
To improve factual consistency and medical relevance, we construct a retrieval corpus $\mathcal{C} = \{c_1, c_2, \dots, c_K\}$ consisting of curated domain-specific resources, including:
(1) general medical knowledge graph, 
(2) clinical guidelines for osteomyelitis treatment and infection management,
(3) evidence-based postoperative outcome studies,
(4) surgical decision support documents,
and (5) institutional recovery protocols and expert narratives.

Given an input patient representation $z$ derived from the previous information extraction module, we encode it using a query encoder $\mathcal{E}_q$ and retrieve the top-$k$ relevant documents using a dense retriever:
\begin{equation}
    \mathcal{R} = \text{Top-}k \left( \text{FAISS} \left( \mathcal{E}_q(z), \{\mathcal{E}_d(c_j)\}_{j=1}^K \right) \right)
\end{equation}
where $\mathcal{E}_d$ denotes the document encoder. The retrieved evidence set $\mathcal{R}$ is then combined with patient-specific prompts for generation.

\noindent \textbf{Expert-Guided Prognostic Evaluation:}
To improve interpretability and anchor the generation process to clinically meaningful concepts, we define a prognostic evaluation $\mathcal{F} = \{f_1, f_2, \dots, f_{12}\}$ comprising twelve expert-selected features related to infection severity, surgical history, and imaging biomarkers. These indicators were identified by orthopedic experts as the most relevant factors for postoperative recovery assessment in chronic osteomyelitis, and are used to guide both structured prompt construction and targeted retrieval. Compared with directly prompting the model using raw multimodal records alone, this expert-guided design provides a more clinically grounded intermediate representation and helps organize heterogeneous patient evidence into a unified prognostic schema.

The twelve indicators span three complementary dimensions and are summarized in Table~\ref{tab:expert_indicators}. Specifically, they include:

\begin{itemize}
    \item \textbf{Infection and Surgical History:} Aetiopathogenesis, Cierny-Mader classification, time intervals between index, debridement, and revision surgeries, prior debridement count, number of interventions between PET-CT and bone reconstruction, implant removal status, and surgical strategy.
    \item \textbf{Clinical Biomarkers:} WBC count.
    \item \textbf{Radiological Indicators:} SUVmax pre/post-debridement, TLG pre/post-debridement, and SUVmax location shift.
\end{itemize}

These indicators guide the construction of structured prompt templates that capture semantically relevant clinical information. These factors jointly capture the major dimensions of postoperative prognosis, including infection origin and severity, surgical burden, systemic inflammatory activity, and residual metabolic evidence from PET-CT. By explicitly encoding these clinically meaningful variables, the framework is able to preserve medically relevant signals even when the underlying records are incomplete, asynchronous, or distributed across different document types.

To make this expert-guided prognostic schema explicit, Table~\ref{tab:expert_indicators} summarizes the twelve indicators used in RAG4Outcome, together with their associated clinical roles and primary source modalities. This design transforms heterogeneous multimodal evidence into a structured and clinically interpretable intermediate representation, improving both prompt consistency and the traceability of the final prognostic prediction.

\begin{table*}[t]
\centering
\caption{Expert-defined prognostic indicators used in RAG4Outcome.}
\label{tab:expert_indicators}
\small
\resizebox{\textwidth}{!}{
\begin{tabular}{l l p{5.5cm} p{4.2cm}}
\hline
\textbf{Category} & \textbf{Indicator} & \textbf{Prognostic relevance} & \textbf{Source modality} \\
\hline
Infection / Surgical History & Aetiopathogenesis & Infection origin and disease mechanism related to recurrence pattern and treatment complexity. & Diagnostic record / EHR \\
Infection / Surgical History & Cierny--Mader classification & Severity stratification based on anatomical type and host condition. & Diagnostic record / EHR \\
Infection / Surgical History & Index-to-debridement interval & Treatment timing signal reflecting disease progression before surgical control. & Surgical timeline \\
Infection / Surgical History & Debridement-to-revision interval & Postoperative evolution and need for additional intervention. & Surgical timeline \\
Infection / Surgical History & Prior debridement count & Surgical burden and chronicity of infection. & Surgical history / EHR \\
Infection / Surgical History & Interventions between PET-CT and reconstruction & Complexity of interim management before definitive reconstruction. & Surgical history / PET-CT timeline \\
Infection / Surgical History & Implant removal status & Whether potentially infection-associated hardware was removed. & Surgical record \\
Infection / Surgical History & Surgical strategy & Overall operative management pathway relevant to expected recovery. & Surgical note \\
Clinical Biomarker & WBC count & Systemic inflammatory/infectious activity marker. & Lab / EHR \\
Radiological Indicator & SUVmax pre-/post-debridement & Peak metabolic activity for residual infection assessment and response evaluation. & PET-CT report \\
Radiological Indicator & TLG pre-/post-debridement & Total metabolic lesion burden related to disease severity. & PET-CT report \\
Radiological Indicator & SUVmax location shift & Spatial change of dominant metabolic focus, indicating persistence or migration of disease. & PET-CT report \\
\hline
\end{tabular}
}
\end{table*}

These indicators guide the construction of structured prompt templates that capture semantically relevant clinical information. In particular, each prognostic factor is instantiated from the extracted patient evidence and used to organize the multimodal record into a form that is more suitable for retrieval and downstream reasoning. This factor-based design also improves robustness in real-world clinical settings, where equivalent prognostic evidence may appear in different formats across PET-CT reports, surgical notes, laboratory tests, and follow-up documentation.

\noindent \textbf{Retrieval-Augmented Generation for Prognostic Prediction:}
The language generation process is handled by a large language model $\mathcal{G}$ (we applied Qwen3-4B here), which takes as input a concatenated prompt comprising patient data summary and retrieved evidence passages:
\begin{equation}
    x = [\text{Prompt}(z); \mathcal{R}]
\end{equation}

The prompt design leverages slot-filling mechanisms, where each prognostic factor $f_i \in \mathcal{F}$ is instantiated with extracted patient data and used to condition retrieval and generation:
\begin{equation}
    \text{Prompt}(z) = \text{TemplateFill}(f_1(z), f_2(z), \dots, f_{12}(z))
\end{equation}
The generation objective is:
\begin{equation}
    y = \mathcal{G}(x) = \arg\max_{y} \prod_{t=1}^{T} P(y_t | y_{<t}, x)
\end{equation}
where $y = \{y_1, y_2, \dots, y_T\}$ is the generated output sequence, consisting of:
(1) a structured textual summary synthesizing patient-specific features, and 
(2) a final outcome prediction label $o \in \{\text{good}, \text{fair}, \text{poor}, \text{very poor}\}$ with supporting rationale.

In our setting, the final prediction is produced in four levels, corresponding to clinically interpretable postoperative outcomes. By combining structured patient-specific factors with retrieved external evidence, the model is encouraged to generate prognostic assessments that are not only coherent and clinically informed, but also more transparent and verifiable.

\subsection{Implementation Details}

In practice, the multimodal inputs are first normalized into textual units that can be consistently consumed by the retrieval and generation modules. PET-CT reports are converted into structured findings with emphasis on lesion activity, anatomical site, and temporal change, while surgical and follow-up records are segmented into clinically meaningful units such as diagnosis, intervention history, laboratory findings, and postoperative evolution. These normalized textual units are then mapped to the predefined prognostic schema and assembled into a patient-specific structured prompt.

For retrieval, we use a dense vector index built from a curated domain-specific corpus containing osteomyelitis-related guidelines, postoperative management literature, recovery assessment documents, and expert-authored reference materials. At inference time, the patient prompt is encoded as a query and matched against the indexed corpus to retrieve the most relevant evidence passages. The retrieved passages are concatenated with the patient summary and then passed to the generator for final prognostic reasoning. This design allows the framework to decouple heterogeneous data processing from downstream outcome prediction, making it more robust to missing modalities and variable documentation quality.

To improve interpretability, the generated output is constrained to include two components: a structured summary of patient-specific risk factors and a final categorical prognosis accompanied by explicit supporting evidence. In this way, the model output remains traceable to both input records and external references, which is essential for high-stakes clinical use.

\section{Experiments and Discussions}
\label{sec:Experiments}

\begin{figure*}[t]
    \centering
    \includegraphics[width=0.85\textwidth]{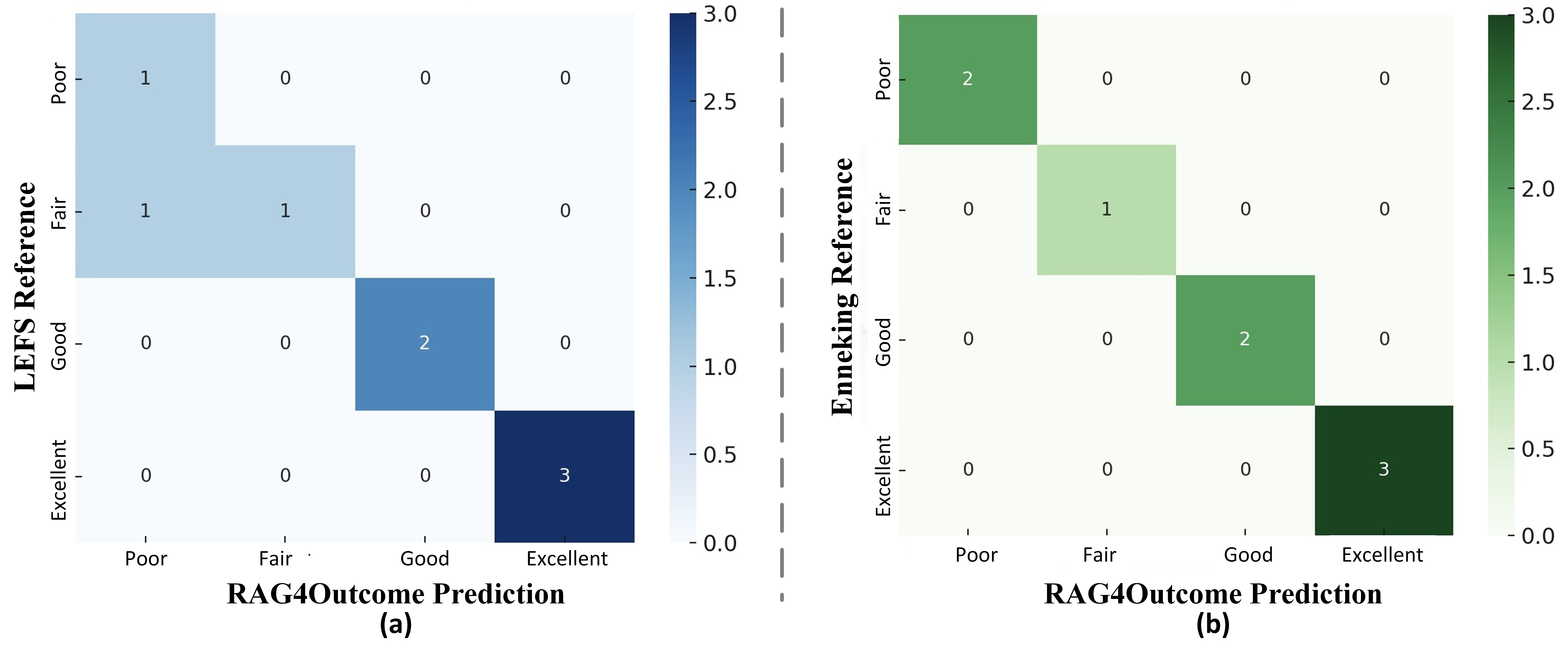}
    \caption{\textbf{Model prediction analysis on 8 case study patients.} (a). Confusion matrix: RAG prediction vs LEFS. (b). Confusion matrix: RAG prediction vs Enneking. }
    \label{fig:Confusion_Matrix}
\end{figure*}

\subsection{Data Collection}
To evaluate the feasibility and interpretability of our proposed RAG4Outcome framework, we conduct preliminary case studies using a real-world dataset collected from a tertiary care hospital in China. The dataset we are constructing is planned to contain clinical information of 230 chronic osteomyelitis patients, each followed up for 3 - 6 years. Each patient case includes multimodal clinical records: PET-CT imaging reports, structured EHR data (diagnoses, surgeries, lab values), and longitudinal follow-up notes. For this workshop submission, we focus on a representative subset of 8 patients, denoted as P1–P8, to demonstrate the effectiveness of our retrieval-augmented prognostic reasoning pipeline. 

The collected cohort reflects the practical complexity of real-world postoperative management. In particular, modality completeness varies substantially across patients, and the temporal spacing between PET-CT examination, surgery, and follow-up is not uniform. Some cases contain rich longitudinal documentation, while others include sparse but clinically critical observations. This heterogeneity makes the cohort suitable for evaluating whether a prognostic framework can remain clinically informative under realistic documentation constraints rather than under artificially standardized multimodal settings.



\subsection{Case Study and Results}

We evaluate the proposed RAG4Outcome framework on eight real-world patient cases drawn from a larger cohort of 230 patients.
The ground-truth outcome severity for each case was determined according to two widely used clinical scoring systems:
\begin{itemize}
\item \textbf{LEFS Score (Lower Extremity Functional Scale):} Outcome levels were categorized based on standard thresholds: \textit{severe functional impairment (0–20)}, \textit{moderate (21–40)}, \textit{mild to moderate (41–60)}, and \textit{mild or normal ($>$60)}.
\item \textbf{Enneking Score:} Outcome Levels were defined as \textit{Poor (0–9)}, \textit{Fair (10–17)}, \textit{Good (18–25)}, and \textit{Excellent ($>$26)}.
\end{itemize}
To ensure consistent evaluation across cases, all outcomes were standardized into a unified four-level taxonomy: \textit{Poor}, \textit{Fair}, \textit{Good}, and \textit{Excellent}. This unified scale follows the Enneking system and serves as the label space for model prediction. 

The eight patient cases were selected to represent a diverse range of outcome severities, clinical presentations, and data completeness levels. Specifically, cases were chosen to cover each of the four unified outcome categories, include both straightforward and clinically challenging instances, and reflect variation in available longitudinal and imaging data. This selection strategy aims to qualitatively demonstrate the model’s reasoning ability and interpretability across different clinical contexts, rather than to establish statistical significance. RAG4Outcome predicts one of these four outcome levels for each case, accompanied by an interpretable explanation grounded in both patient-specific features and retrieved clinical evidence. Model confidence was quantified using the softmax-normalized log-likelihood over the output distribution.

We visualize and analyze model performance from three perspectives: agreement with clinical reference labels, confidence score distribution, and inter-method prediction alignment. Results are shown in Figure~\ref{fig:Confusion_Matrix}.

\begin{figure*}[t]
    \centering
    \includegraphics[width=0.95\textwidth]{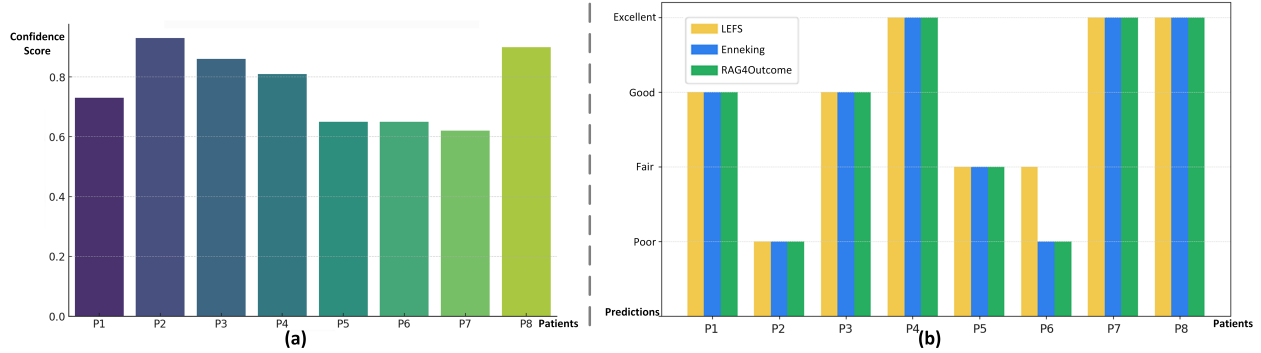}
    \caption{\textbf{Model prediction analysis on 8 case study patients.} (a). Confidence scores per patient. (b). Comparison of LEFS, Enneking, and RAG predictions for each patient.}
    \label{fig:Results_and_Comparison}
\end{figure*}

\noindent\textbf{Prediction Agreement with Clinical Scores.}  
Figures~\ref{fig:Confusion_Matrix} (a) and (b) show confusion matrices comparing RAG4Outcome predictions with LEFS and Enneking ground truth labels, respectively. Most predictions fall on the diagonal, indicating strong agreement. A few off-diagonal entries, such as between “poor” and “fair” in Figure~\ref{fig:Confusion_Matrix}(a), suggest minor mismatches likely due to subjective boundaries in clinical scoring. This underscores the robustness of our model across different reference systems.

\noindent\textbf{Confidence Score Evaluation.}  
The confidence scores range from 0.62 to 0.92, as shown in Figure~\ref{fig:Results_and_Comparison}(a), indicating generally high model certainty. Lower confidence in patients P5–P7 corresponds to ambiguous conditions, which also fall near the clinical decision boundaries in LEFS and Enneking scoring.

\noindent\textbf{Three-Way Prediction Comparison.}  
Figure~\ref{fig:Results_and_Comparison}(b) presents a side-by-side comparison of LEFS, Enneking, and RAG4Outcome predictions. Most patients show consistent results across all three methods. Discrepancies observed in P5 highlight the potential advantage of retrieval-augmented reasoning in resolving subjective ambiguities in clinical assessment.

Qualitative inspection of the retrieved evidence further supports the validity of the proposed framework. Across the selected cases, the retrieved passages frequently focused on clinically meaningful concepts such as recurrence risk after repeated debridement, interpretation of postoperative inflammatory activity, and the prognostic relevance of PET-CT metabolic indicators. In cases with clearer recovery trajectories, the retrieved evidence was highly consistent with the final prediction and the clinical scores. In more ambiguous cases, retrieval provided additional contextual support that helped the model articulate why a borderline case should be assigned to a more conservative or more favorable outcome category. This suggests that the value of RAG4Outcome lies not only in final label prediction, but also in structuring clinically relevant evidence for transparent decision support.

\subsection{Ablation Study}

While the case study analysis highlights the interpretability of the proposed framework, we further perform a small-scale ablation study to better understand the contribution of retrieval augmentation and PET-CT-derived evidence to the final prognostic prediction.
To further examine the contribution of each core component in RAG4Outcome, we conducted an ablation study on the eight patient cases used in our qualitative evaluation. Specifically, we compared three model configurations: (1) the full model, which uses multimodal patient information together with retrieval-augmented external medical evidence; (2) a variant without RAG, which disables retrieval from the external knowledge corpus and relies solely on patient-specific clinical information; and (3) a variant without PET-CT-derived evidence, which uses EHR-based clinical records together with the RAG module but excludes radiology-related postoperative findings.

\begin{table*}[t]
\centering
\caption{Ablation study of different RAG4Outcome variants on the 8-case evaluation subset.}
\label{tab:ablation}
\resizebox{\linewidth}{!}{
\begin{tabular}{l c p{9.3cm}}
\hline
\textbf{Variant} & \textbf{Macro-F1} & \textbf{Interpretation} \\
\hline
RAG4Outcome (Full) & 0.875 & Best overall performance, showing that multimodal evidence and retrieval augmentation jointly improve prognostic consistency. \\
RAG4Outcome w/o Retrieval & 0.625 & Removing the retrieval module substantially weakens performance, indicating that external domain knowledge helps resolve uncertainty in patient-specific evidence. \\
RAG4Outcome w/o PET-CT-derived evidence & 0.750 & Excluding PET-CT-related findings reduces performance, suggesting that radiology-informed postoperative cues provide important complementary prognostic signals. \\
\hline
\end{tabular}
}
\end{table*}

Given the small cohort size and the four-category prognostic setting, we adopted the Macro-F1 score to evaluate the agreement between model predictions and the unified reference labels (\textit{Poor}, \textit{Fair}, \textit{Good}, and \textit{Excellent}). As shown in Table~\ref{tab:ablation}, the full model achieved the best performance, indicating the benefit of combining multimodal clinical evidence with retrieval-augmented reasoning. When the RAG module was removed, performance dropped substantially, suggesting that external medical knowledge plays an important role in supporting robust decision-making, particularly for clinically ambiguous cases. Removing PET-CT-derived evidence also led to a noticeable decline, indicating that radiology-informed postoperative cues remain important for accurate prognosis assessment.

Overall, the ablation results support the design motivation of RAG4Outcome. Both multimodal evidence integration and retrieval augmentation contribute to model effectiveness, while the combination of the two produces predictions that are more consistent with the unified clinical reference. These findings further suggest that external retrieval helps compensate for uncertainty and incompleteness in patient records, whereas PET-CT-derived indicators provide additional prognostic signals that may not be fully captured from textual EHR evidence alone. We note that this ablation study is conducted on a small case subset and should therefore be interpreted as preliminary evidence of component-wise contribution.

%

\subsection{Qualitative Analysis}

To further illustrate the interpretable reasoning process of RAG4Outcome, we present two representative patient cases with different prognostic characteristics. Rather than focusing only on the final predicted label, this analysis highlights how multimodal patient evidence and retrieved external knowledge are jointly organized into a clinically meaningful reasoning pathway.

\noindent\textbf{Case P2 (Excellent Outcome).}
This patient presented with a relatively localized infection pattern and underwent timely surgical intervention. PET-CT-derived evidence indicated a focal metabolic signal with limited spread, while the clinical record suggested a favorable surgical trajectory. The model retrieved external evidence related to early intervention and postoperative infection control, and finally predicted an \textit{Excellent} outcome. The generated rationale emphasized localized disease burden, timely treatment, and the absence of strong indicators of persistent postoperative activity. This prediction was consistent with the long-term follow-up outcome showing stable recovery without recurrence.

\noindent\textbf{Case P6 (Fair/Poor Boundary).}
In contrast, this patient exhibited a more complex clinical course, including multiple prior debridement procedures and less favorable postoperative characteristics. PET-CT-derived evidence suggested more diffuse metabolic activity, and the EHR record reflected a higher surgical burden. These signals alone could support a pessimistic interpretation. However, after retrieval augmentation, the model incorporated additional external evidence related to host condition and treatment strategy, and produced a \textit{Fair} prediction rather than a strictly \textit{Poor} one. The generated rationale noted a guarded prognosis together with the possibility of stabilization under appropriate management. 

Overall, these representative cases show that the value of RAG4Outcome lies not only in categorical prediction, but also in its ability to expose an interpretable reasoning path from multimodal records to retrieved evidence and final clinical judgment.

\subsection{Discussion}

The results suggest that \textbf{RAG4Outcome} provides consistent and clinically reliable predictions. Compared to black-box classifiers, it offers structured justifications aligned with clinical reasoning. Retrieved evidence often refers to literature on infection recurrence, surgical timing, or imaging biomarkers, providing grounded support for its predictions. We observed high consistency between the model’s generated rationale and physician assessments. The agreement with expert scoring systems, combined with the ability to explain its decisions using external references, supports the model’s practical value for prognostic prediction in chronic osteomyelitis.
We acknowledge that the evaluation on eight cases is preliminary and does not yet allow conclusions about generalizability or robustness.
The primary goal of this case study is to illustrate the model’s interpretability and clinical reasoning process, not to provide statistically significant validation.
A larger-scale quantitative evaluation involving the full 230-patient cohort (and potentially additional cases) is planned for future work to further assess model performance, generalizability, and data-driven variability.
Nevertheless, the selected cases provide an informative cross-section of clinical diversity within the broader cohort, demonstrating how RAG4Outcome handles both typical and complex scenarios.

\section{Conclusion}
\label{sec:conclusion}
In this work, we present RAG4Outcome, a retrieval-augmented multimodal framework for postoperative outcome prediction in chronic osteomyelitis. Our method integrates heterogeneous clinical data into a unified reasoning process supported by expert-defined prognostic indicators and external clinical knowledge. Case study results demonstrate promising consistency with gold-standard clinical scoring systems, alongside high interpretability and confidence transparency.
For future work, we plan to expand the evaluation to the full patient cohort. Moreover, we aim to refine the retrieval corpus and explore model alignment techniques to enhance factual consistency and domain safety in broader clinical scenarios.

\section*{Acknowledgments}
This work was supported by the Pudong New Area Science and Technology Development Fund-Public Institution Livelihood Research Special Project (Healthcare) (No. PKJ2024-Y05), and the Clinical Research Program funded by Shanghai Sixth People’s Hospital Affiliated to Shanghai Jiao Tong University School of Medicine (No. ynts202201). This work was also conducted in collaboration with LinkIntelli Technology.

{
    \small
    \bibliographystyle{ieeenat_fullname}
    \bibliography{main}
}


\end{document}